\documentclass[preprint,aps,showpacs,preprintnumbers,nofootinbib,eqsecnum]{revtex4}

\usepackage{amsmath}
\usepackage{amssymb}
\usepackage{latexsym}
\usepackage{graphicx}
\usepackage{dcolumn}
\usepackage[tight]{subfigure}

\newcommand{\tr}{{\rm tr}}
\newcommand{\SO}{{\it SO}}
\newcommand{\SU}{{\it SU}}
\newcommand{\U}{{\it U}}

\begin{document}
%----------------------------------------------------------------

%================================================================
%\date{\today}
%================================================================

\preprint{RIKEN-TH-75}
\preprint{hep-th/0608031}

%================================================================

\title{%
Dynamical Generation of Non-Abelian Gauge Group \\
via the Improved Perturbation Theory}

%----------------------------------------------------------------

\author{T.~Aoyama}
%\email{}
\affiliation{Theoretical Physics Laboratory, RIKEN, Wako, 351-0198, Japan }

\author{T.~Kuroki}
%\email{}
\affiliation{Theoretical Physics Laboratory, RIKEN, Wako, 351-0198, Japan }

\author{Y.~Shibusa}
%\email{}
\affiliation{Theoretical Physics Laboratory, RIKEN, Wako, 351-0198, Japan }

%================================================================
\begin{abstract}
It was suggested that 
the massive Yang-Mills-Chern-Simons matrix model has three phases 
and that in one of them a non-Abelian gauge symmetry 
is dynamically generated. 
The analysis was at the one-loop level around a classical solution 
of fuzzy sphere type. 
We obtain evidences that three phases are indeed realized 
as nonperturbative vacua by using the improved perturbation theory. 
It also gives a good example 
that even if we start from a trivial vacuum, 
the improved perturbation theory around it enables us 
to observe nontrivial vacua.    
\end{abstract}
%================================================================

%----------------------------------------------------------------
\pacs{ 11.25.-w, 11.25.Yb, 11.30.Qc, 02.30.Mv }

% 11.25.-w Strings and branes
% 11.25.Yb M theory
% 11.30.Qc Spontaneous and radiative symmetry breaking
% 02.30.Mv Approximations and expansions

%----------------------------------------------------------------

\maketitle
%================================================================

%%%%%%%%%%%%%%%%%%%%%%%%%%%%%%%%%%%%%%%%%%%%%%%%%%%%%%%%%%%%%%%%%
\section{Introduction}
%%%%%%%%%%%%%%%%%%%%%%%%%%%%%%%%%%%%%%%%%%%%%%%%%%%%%%%%%%%%%%%%%
\label{sec:intro}

It is now widely believed that some gauge theories or matrix models 
would provide nonperturbative formulations of string theories 
in the large-$N$ limit 
\cite{Banks:1996vh,Ishibashi:1996xs,Maldacena:1997re}. 
In such approaches, it is quite important 
to clarify how they quantize gravity, namely, 
how they reconcile the general relativity and quantum theory. 
It is also of great importance to confirm 
whether at a low energy regime 
they can be effectively described by quantum field theories 
of gauge fields with the non-Abelian gauge group 
of the standard model type. 
However, in the context of the matrix models it is not yet known 
how the gauge group of the standard model is dynamically generated. 
It is natural that it is embedded in the original $\U(N)$ gauge group 
of a matrix model we start from and that it emerges dynamically 
in the large-$N$ limit 
\cite{Iso:1999xs} 
via the Higgs mechanism or the confinement, for example. 
Therefore, it is interesting and necessary to construct 
matrix models that incorporate such a mechanism 
and to study them in a nonperturbative manner. 

As a first step toward this program, the bosonic 
massive Yang-Mills-Chern-Simons (YMCS) matrix model 
was analyzed in 
Ref.~\cite{Azuma:2005bj}, 
which is obtained by the large-$N$ reduction 
\cite{Kawai:1982nm} 
of the $\U(N)$ Yang-Mills theory with the Chern-Simons term 
in three dimensions and by adding the mass term. 
Without the mass term, it was shown in 
Ref.~\cite{Azuma:2004zq} 
that the YMCS matrix model has 
two phases, namely, the Yang-Mills phase and the fuzzy sphere phase. 
In the former phase observables show qualitatively the same behavior 
as those in the pure Yang-Mills model without the Chern-Simons term, 
while in the latter a single fuzzy sphere is a stable and dominant 
configuration due to the Chern-Simons term.  
It was claimed in 
Ref.~\cite{Azuma:2005bj} 
that if the mass term is present 
there appears an interesting third phase (``5-brane phase'') 
in addition to these two phases, 
in which a dominant configuration is expected to be  
$\mathcal{O}(N)$ coincident fuzzy spheres of small size 
and thus a non-Abelian gauge group with rank of $\mathcal{O}(N)$ 
is dynamically generated as a subgroup 
of the original $\U(N)$ gauge group
\cite{Iso:2001mg}. 
Although in the fuzzy sphere phase 
comparison between the one-loop calculation 
around a single fuzzy sphere configuration and the Monte Carlo data 
shows one-loop dominance, namely higher-loop corrections 
are suppressed in the large-$N$ limit
as long as 1PI diagrams are concerned \cite{Azuma:2004zq}, 
it is not \textit{a priori} trivial 
whether the analysis at the one-loop level 
is sufficient even in the presence of mass term. 

In order to examine whether the 5-brane phase found in 
Ref.~\cite{Azuma:2005bj} 
is indeed realized as a nonperturbative vacuum,  
we have to investigate 
the massive YMCS matrix model in a nonperturbative method. 
Toward this aim, we employ here the improved perturbation theory 
developed in Ref.~\cite{Kawai:2002jk}, 
which is motivated by 
the improved mean field approximation (IMFA). 
This concept is also called the Gaussian approximation or 
Gaussian expansion method 
and was introduced to a large-$N$ reduced model in 
Ref.~\cite{Kabat:1999hp,Oda:2000im}, 
which has been further investigated in various manners 
\cite{Nishimura:2001sx,Kawai:2002ub,Nishimura:2002va}. 
This technique is a sort of variational method 
\cite{Stevenson:1981vj} 
to extract nonperturbative information on the vacuum of a theory. 
The IMFA is indeed reduced to the standard mean field approximation 
at the leading order, 
and the inclusion of higher-order contributions 
may be viewed as systematic improvement of the approximation. 
The IMFA has been applied successfully to a number of models 
to yield nonperturbative results  
\cite{Kawai:2002jk,Kabat:1999hp,Oda:2000im,Nishimura:2001sx,Kawai:2002ub,Nishimura:2002va,Zhong:2003xr,Kawamoto:2003kn,Aoyama:2006di,Aoyama:2006je,Aoyama:2005nd}. 
In particular, the IMFA was applied to the IIB matrix model 
first in 
Ref.~\cite{Nishimura:2001sx} 
and provided the interesting nonperturbative observation 
that our four-dimensional space-time is 
realized in the IIB matrix model as 
the anisotropy of eigenvalue distribution 
of ten bosonic Hermitian matrices. 
This result was further confirmed in 
Refs.~\cite{Kawai:2002jk,Kawai:2002ub,Aoyama:2006di,Aoyama:2006je}. 

The improved perturbation theory was developed as such 
a reformulation of the IMFA that it is considered as 
a reorganization of a perturbative series. 
It opened a way to taking account of more general types 
of action than that had been treated in the IMFA 
or the Gaussian expansion method. 
Thus in the present article we examine the matrix model 
with Chern-Simons term that has a cubic interaction term 
in the framework of the improved perturbation theory. 
We expect that the observations obtained in this work would be 
instructive to future studies of more nontrivial and interesting 
nonperturbative phenomena 
via the improved perturbation theory, 
in particular, dynamical generation of gauge symmetry 
in the IIB matrix model. 

In the prescription of the IMFA and the improved perturbation theory, 
a set of artificial parameters are introduced through a nominal deformation 
in the original action of the model. 
It leads to a sort of consistency conditions for these parameters 
that the original theory should not depend on them. 
In the space of these parameters, the conditions may be realized 
as a region in which physical quantities are least sensitive to 
the variation of the parameters 
\cite{Stevenson:1981vj}. 
Such a region is called ``plateau''.  
As stressed in 
Refs.~\cite{Kawai:2002jk,Kawai:2002ub}, 
formation of a plateau provides an essential criterion 
to see whether or not the improved perturbation theory works. 
It is shown in 
Ref.~\cite{Aoyama:2005nd}
that in a system which exhibits a phase transition, 
there may appear several plateaux. 
Therefore, it is interesting to discuss the phase structure of 
the massive YMCS matrix model in the framework 
of the improved perturbation theory through the patterns of 
emergence of plateaux that correspond to various phases. 

Another motivation of applying the improved perturbation theory 
to the massive YMCS matrix model comes from the fact that 
even if we start from the perturbative vacuum, 
the improved perturbation theory can provide information 
on a nonperturbative vacuum which does not always correspond 
to a classical configuration. 
Therefore, in principle, 
even if we start from the trivial vacuum which corresponds 
to the Yang-Mills phase, 
we would be able to identify 
the fuzzy sphere phase and the 5-brane phase 
when they are realized as nonperturbative vacua. 
We will see that this is indeed the case. 
This is in contrast with previous analyses of fuzzy sphere type 
configurations 
based on the perturbation theory around them 
\cite{Azuma:2005bj,Azuma:2004zq,Imai:2003vr,Azuma:2004yg}. 

The organization of this paper is as follows: in the next section 
we review the massive YMCS matrix model and its three phases. 
In Section~\ref{sec:improve}, 
we also give a review of the improved perturbation theory, 
which is applied to the massive YMCS matrix model 
in Section~\ref{sec:mymcsmm}. 
The results are presented 
in Section~\ref{sec:result}. 
Section~\ref{sec:discussion} 
is devoted to discussions. The tables 
are put collectively at the 
end of the paper.

%%%%%%%%%%%%%%%%%%%%%%%%%%%%%%%%%%%%%%%%%%%%%%%%%%%%%%%%%%%%%%%%%
\section{The model}
%%%%%%%%%%%%%%%%%%%%%%%%%%%%%%%%%%%%%%%%%%%%%%%%%%%%%%%%%%%%%%%%%
\label{sec:model}

The massive Yang-Mills-Chern-Simons matrix model is defined 
by the action 
\begin{equation}
	S = 
	N\tr\left(-\frac{1}{4}[A_{\mu},A_{\nu}]^2
	+\frac{i}{3}\alpha\epsilon_{\mu\nu\lambda}A_{\mu}A_{\nu}A_{\lambda}
	+\frac{1}{2}m^2A_{\mu}^2\right),
\label{eq:action}
\end{equation}
where $A_{\mu}$ ($\mu=1\sim 3$) are $N\times N$ Hermitian matrices. 
We will consider this model in the large-$N$ limit. 
We can obtain this model by adding the mass term 
to the large-$N$ reduced model of the $\U(N)$ Yang-Mills 
theory with the Chern-Simon term in three dimensions. 
This model has two parameters $\alpha$, $m$ and 
we are interested in the free energy of this model 
\begin{equation}
	F=-\frac{1}{N^2}\log Z,
	\qquad
	Z=\int dA_{\mu}\, e^{-S},
\label{free energy}
\end{equation}
as a function of them, by which we can determine 
the phase structure of this model. 

The equation of motion derived from Eq.~(\ref{eq:action}) is 
\begin{equation}
	[A_{\lambda},[A_{\lambda},A_{\mu}]]
	+\frac{i}{2}\alpha\epsilon_{\mu\nu\lambda}[A_{\nu},A_{\lambda}]
	+m^2A_{\mu}=0.
\end{equation}
Apart from a trivial solution $A_{\mu}=0$, 
this model admits a classical solution of fuzzy sphere type given by 
\begin{equation}
	A_{\mu}=\chi L_{\mu},
	\qquad
	\mu=1\sim 3, 
\end{equation}
where $L_{\mu}$'s is the $N$-dimensional representation 
of $\SU(2)$ algebra satisfying 
\begin{equation}
	[L_{\mu},L_{\nu}]=i\epsilon_{\mu\nu\lambda}L_{\lambda},
\label{su(2)}
\end{equation}
and $\chi=(\alpha+\sqrt{\alpha^2-8m^2})/4$. 
If $m^2>\alpha^2/8$, this type of solution ceases to exist. 
Generically $L_{\mu}$ can be a 
reducible representation of the $\SU(2)$ algebra 
which consists of the $n_i$-dimensional irreducible representations 
for $i=1\sim s$, where $\sum_{i=1}^sn_i=N$. 
In this case without loss of generality $A_{\mu}$ can be brought 
into the form 
\begin{equation}
	A_{\mu}=\chi
	\begin{pmatrix}
		L_{\mu}^{(n_1)} & & & \\
		 & L_{\mu}^{(n_2)} & & \\
		 & & \ddots & \\
		 & & & L_{\mu}^{(n_s)} \\
	\end{pmatrix},
\end{equation}
where $L_{\mu}^{(n)}$ denotes the $n$-dimensional 
irreducible representation. 
Substituting this into the action (\ref{eq:action}), 
we obtain the classical action of this solution as 
\begin{equation}
	S_{\text{cl}} = 
	Nf(\chi)\sum_{i=1}^s\frac{n_i(n_i^2-1)}{4},
\label{classical action}
\end{equation}
where $f(\chi)=\chi^4/2-\alpha\chi^3/3+m^2\chi^2/2$. 
When $\alpha^2>9m^2$, $f(\chi)$ is always negative, thus 
Eq.~(\ref{classical action}) implies that a single fuzzy sphere 
($s=1$, $n_1=N$) is the most dominant configuration 
as long as $\alpha^2$ and $m^2$ are not so small 
that $S_{\text{cl}} \sim \mathcal{O}(1)$. 
Then the system is in the fuzzy sphere phase. 
On the other hand, when $\alpha^2<9m^2$, $f(\chi)>0$ and 
the trivial solution $A_{\mu}=0$ becomes stable. 
Moreover, whenever $\alpha$ is so small, 
the one-loop analysis around a background 
\begin{equation}
	A_{\mu} = 
	\text{diag}(x_{\mu}^{(1)},x_{\mu}^{(2)},\cdots,x_{\mu}^{(N)}),
\label{trivial solution}
\end{equation}
shows that in general there is an attractive force 
between the eigenvalues of $A_{\mu}$, 
and it makes their distribution shrink until the perturbative 
calculation becomes no longer valid 
\cite{Hotta:1998en}. 
Then they form a ``solid ball'' and the system is in the Yang-Mills phase. 
However, as shown in 
Ref.~\cite{Azuma:2005bj}, 
this is not the end of the story. 
So far we have discussed the phase of the model 
based on the classical action (\ref{classical action}). 
If we take account of the one-loop contribution to the free energy, 
we can find the third phase for the moderate $\alpha$ and $m$ 
satisfying $8m^2<\alpha^2<9m^2$ in which $\mathcal{O}(N)$ copies 
of the small fuzzy sphere appear as the true vacuum. 
This phase is called ``5-brane phase'' in 
Ref.~\cite{Azuma:2005bj} 
because this configuration is analogous to the one 
that is interpreted 
\cite{Maldacena:2002rb} 
as coinciding transverse 5-branes in the BMN matrix model 
\cite{Berenstein:2002jq}. By considering configurations 
with fuzzy spheres of various sizes, the phase diagram 
was proposed in Ref.~\cite{Azuma:2005bj} 
at the one-loop level.\footnote
{$\alpha$ and $\rho$ in Ref.~\cite{Azuma:2005bj} are written 
as $\alpha/2$ and $2m/\alpha$ respectively by using 
the parameters in Eq.~(\ref{eq:action}).}
The one-loop calculation is reliable 
in some parameter regions 
provided that the perturbative series around classical configurations 
are well-defined.\footnote
{In particular, the presence of the 5-brane phase 
is shown in Ref.~\cite{Azuma:2005bj} in the region $\alpha\gg 1$ 
where the one-loop calculation is justified.}

However, in a general parameter region it is not clear 
that the one-loop calculation is sufficient and even if so, 
the perturbative series would be asymptotic. 
Therefore, it is interesting to confirm that 
the phase diagram in Ref.~\cite{Azuma:2005bj} is true 
even at a nonperturbative level 
by a totally different method. As such, we employ 
the improved perturbation theory developed in 
Ref.~\cite{Kawai:2002jk}, which works well 
even for some asymptotic series.

%%%%%%%%%%%%%%%%%%%%%%%%%%%%%%%%%%%%%%%%%%%%%%%%%%%%%%%%%%%%%%%%%
\section{The improved perturbation theory}
%%%%%%%%%%%%%%%%%%%%%%%%%%%%%%%%%%%%%%%%%%%%%%%%%%%%%%%%%%%%%%%%%
\label{sec:improve}

In order to describe the idea of the improved perturbation theory, 
we consider as an illustration a one-matrix model 
defined by the action
\begin{equation}
	S = \frac{N}{4}\tr \phi^4,
\label{one-matrix model}
\end{equation}
where $\phi$ is an $N\times N$ Hermitian matrix. 
Suppose we are interested in the free energy 
\begin{equation}
	N^2F = -\log \int d\phi \exp\left(-\frac{N}{4}\tr \phi^4\right),
\label{F of one-matrix model}
\end{equation}
which cannot be computed by the standard perturbation theory 
for lack of the mass term (quadratic term) in Eq.~(\ref{one-matrix model}). 
Therefore, let us formally add and subtract the mass term in the action, 
one of which is regarded as a perturbation. 
Then we introduce a formal coupling constant $g$ and construct 
a perturbative series in terms of $g$. 
$g$ will be taken to be 1 in the end: 
\begin{equation}
	N^2F = \left.-\log \int d\phi 
	\exp\left(-N\tr\left(\frac{g}{4}\phi^4-g\frac{m_0^2}{2}\phi^2
	+\frac{m_0^2}{2}\phi^2\right)\right)\right|_{g=1}, 
\label{add and subtract}             
\end{equation} 
where $m_0$ is an artificially introduced parameter. 
It can be restated in the following way. 
If we define the free energy of a massive theory as
\begin{equation}
	N^2F(m^2) = 
	-\log \int d\phi 
	\exp\left(
		-N\tr\left(\frac{g}{4}\phi^4+\frac{m^2}{2}\phi^2\right)
	\right),
\end{equation}
the prescription above is equivalent to  
calculating the free energy 
of the massless theory (\ref{F of one-matrix model}) 
from the massive theory with mass parameter replaced as 
\begin{equation}
	F(0)\bigl|_{g=1} = F(m_0^2-gm_0^2)\bigr|_{g=1}.
\label{eq:F_by_improve}
\end{equation}
The perturbative series in terms of $g$ obtained from the RHS 
of Eq.~(\ref{eq:F_by_improve}) 
is called the improved perturbative series. 
Note that if we can sum up this series at the full order, 
the free energy is completely independent of $m_0$ which 
is introduced artificially 
in order to make the perturbation theory feasible. 
 
In practice, we can calculate the RHS in Eq.~(\ref{eq:F_by_improve}) 
only up to a finite order. Then the dependence on $m_0$ emerges 
in the improved perturbative series. In order to obtain the exact value 
of the free energy, we need to determine the value of $m_0$ somehow. 
Here we adopt the principle of \textit{minimal sensitivity} 
\cite{Stevenson:1981vj} 
as a guiding principle: 
the improved series should depend least on $m_0$. 
It is because the original theory does not depend on $m_0$ 
which was introduced through a nominal shift of parameter. 
If there exists a region in the parameter space of $m_0$ 
where the dependence on $m_0$ vanishes effectively, 
the \textit{exact} value would be reproduced there. 
We call such a region as ``plateau''. 
The principle of minimal sensitivity is realized as 
the emergence of plateau, in which the improved series 
stays stable against any variation of the artificial parameter. 
 
As for the one-matrix model (\ref{F of one-matrix model}), 
we can identify a clear plateau even up to the 8th order, 
and there the improved perturbative series 
reproduces the exact value of the free energy 
with more than 99.5\% accuracy
\cite{Kawai:2002ub}. 
It is a striking result that although 
the original perturbative series of the massive theory 
has a radius of convergence $g/m^4 < 1/12$ 
\cite{Brezin:1977sv}, 
we have obtained a good approximate value at $m=0$ 
by this technique. 
The plausibility and applicability of the improved 
perturbation theory is not yet clear in the mathematically 
strict sense. Nevertheless 
it has been successfully used in various models. 
Some examples are also given in 
Ref.~\cite{Kawai:2002jk}. 

Note that at the first order in $g$ 
if we require $\partial F_1^{\text{imp}}/\partial m_0=0$ 
as a ``plateau'' condition, this reduces to the self-consistency 
condition in the standard mean field approximation. 
In this sense, the improved perturbation 
series is a natural generalization of the mean field approximation. 

The scheme presented so far can be generalized 
to a generic series that may have more than one parameter. 
Thus we arrive at the concept of 
the improved Taylor expansion 
\cite{Kawai:2002jk} 
as follows. 
Let us assume that an observable of a theory would be exactly 
described by a function $F(\lambda, \xi)$. 
Here $\lambda$ is a coupling constant and $\xi$ 
collectively represents parameters of the model such as a mass. 
Perturbation theory provides an expansion of $F$ as a power 
series of $\lambda$ about $\lambda = 0$, 
with $n$th coefficient denoted as $f_n(\xi)$: 
\begin{equation}
	F(\lambda, \xi) 
	= \sum_{n=0}^{\infty}\,\lambda^n\,f_{n}(\xi) \,.
\label{eq:taylor_series}
\end{equation}
Next we consider a modification of the series according to the following 
prescriptions. First we perform a shift of parameters:
\begin{align}
	\lambda &\longrightarrow g\,\lambda \,, \nonumber \\
	\xi     &\longrightarrow \xi_0 + g (\xi - \xi_0) ,
\label{eq:improve_parameter}
\end{align}
where we have introduced $g$ as a formal expansion parameter, 
and $\xi_0$ as a set of artificial parameters. 
We deform the series by the substitution (\ref{eq:improve_parameter}), 
and then we reorganize the series in terms of $g$ 
up to $g^k$ and drop the $\mathcal{O}(g^{k+1})$ terms, 
and finally set $g$ to 1. 
Thus we obtain the improved perturbative series $F^\text{imp}_k$ as
\begin{equation}
	F(\lambda, \xi) \,
	\longrightarrow \,
	F^\text{imp}_k (\lambda, \xi; \xi_0)
	=
	\left.F(g\lambda, \xi_0+g(\xi-\xi_0))\right|_{\text{up
	to}~\mathcal{O}(g^k),~g=1}.
\label{improved series}
\end{equation}
Here, a notation $\bigr|_{\text{up to}~\mathcal{O}(g^k),~g=1}$ 
represents the operation that 
we disregard the $\mathcal{O}(g^{k+1})$ terms and then put $g$ to 1. 

More concretely, the improved perturbative series up to 
$k$th order $F^\text{imp}_k$ is made by the procedure: 
\begin{eqnarray}
\label{shibusa}
	F_k(\xi)
	& \equiv &
	\sum_{n=0}^{k}\,\lambda^n\,f_{n}(\xi)
	\\
\label{eq:step1}
	& \longrightarrow &
	\sum_{n=0}^{k}\,(g\lambda)^n\,f_n( \xi_0+g(\xi-\xi_0) )
	\\
\label{eq:step2}
	& = &
	\sum_{\ell=0}^{\infty}\,g^\ell\,
	\sum_{n=0}^{\min(k,\ell)}\,\lambda^n\,
	\frac{1}{(\ell-n)!}\,(\xi-\xi_0)^{\ell-n}\,f_n^{(\ell-n)}(\xi_0)
	\\
\label{eq:step3}
	& \longrightarrow &
	\sum_{\ell=0}^{k}\,g^\ell\,
	\sum_{n=0}^{\ell}\,\lambda^n\,
	\frac{1}{(\ell-n)!}\,(\xi-\xi_0)^{\ell-n}\,f_n^{(\ell-n)}(\xi_0)
	\\
\label{eq:step4}
	& \longrightarrow &
	\sum_{n=0}^{k}\,\lambda^n\,
	\sum_{\ell=0}^{k-n}\,
	\frac{1}{\ell!}\,(\xi-\xi_0)^\ell\,f_n^{(\ell)}(\xi_0)
	\\
\label{eq:defofimp}
	& \equiv & 
	F^\text{imp}_k(\lambda,\xi;\xi_0)
	\,.
\end{eqnarray}
Here, $f_n^{(\ell)}$ is the $\ell$th derivative of $f_n(\xi)$ 
with respect to $\xi$.
At first we apply the transformation (\ref{eq:improve_parameter}) to 
the original $k$th order perturbative series in Eq.~(\ref{eq:step1}).
Next the series is then reorganized in terms of $g$ 
in Eq.~(\ref{eq:step2}).
At this stage, we must take into account the extra $g$-dependence as 
$\xi_0+g(\xi-\xi_0)$ in the coefficient $f_n$.  
Then we drop the $\mathcal{O}(g^{k+1})$ terms in Eq.~(\ref{eq:step3}), 
and finally set $g$ to 1 in Eq.~(\ref{eq:step4}). 
In this way we obtain the improved series $F^\text{imp}_k$ 
(\ref{eq:defofimp}).

It turns out that 
by the procedure of the improved perturbation theory, 
each coefficient of the original series $f_n(\xi)$ are 
replaced by a particular combination of the coefficients 
in its Taylor expansion 
about a shifted point $\xi_0$, 
\begin{equation}
	f_n(\xi) 
	\longrightarrow 
	\widetilde{f_n}(\xi; \xi_0) = 
	\sum_{\ell=0}^{k-n}\,
		\frac{1}{\ell!}\,(\xi-\xi_0)^\ell\,f_n^{(\ell)}(\xi_0).
\label{eq:deformed_coeff}
\end{equation}
{}From this expression, we immediately find that even if 
$F_k(\xi)$ in Eq.~(\ref{shibusa}) shows worse behavior, 
the improved perturbative series $F_k^\text{imp}(\xi;\xi_0)$ 
in Eq.~(\ref{eq:defofimp}) often becomes mild as a function 
of $\xi$. 

In this procedure, 
the foregoing example of one matrix model corresponds to a case  
of $\lambda=1$ and $\xi\equiv m^2=0$ where 
the improved perturbative series $F^\text{imp}_k$ boils down to the expression (\ref{eq:F_by_improve}).

Because the emergence of a plateau is an essential 
criterion to see whether the improved perturbation theory works well, 
a main task is identification of plateau with respect to 
the artificial parameter $\xi_0$ at each value of the parameter $\xi$. 
Although we do not yet have a rigorous definition of plateaux, 
we could identify them by 
their properties investigated in 
\cite{Kawai:2002jk,Kawai:2002ub,Aoyama:2005nd,Aoyama:2006je} as follows. 
The ideal realization of the plateau 
may have such a property that the improved series is totally independent 
of the artificial parameters $\xi_0$ in a certain region. 
It involves a situation that 
all orders of derivatives of the improved series 
with respect to $\xi_0$ are zero in that region. 
However, such an ideal plateau is not realized in practical cases 
because we have series only of finite order. 
Typical profile of the improved series that forms plateau 
exhibits a flat region in which the series fluctuates bit by bit; 
it would consist of a number of local maxima and minima. 
Thus, we consider the accumulations of extrema 
as indications of plateau (or its candidates).

In some cases it occurs that 
there is a region in which 
$F^\text{imp}_k$ becomes stable but varies gently without 
forming extrema. 
We should also take into account this region as a plateau. 
However, we have to mention that we often exclude such an asymptotic 
behavior that the series becomes flat at large values of 
parameters $\xi_0$.

For each value of $\xi$, there may appear more than one plateau, 
each of which corresponds 
to either a stable physical state or an unstable or metastable state 
(local minimum). 
By comparing the values of the improved free energy at these plateaux 
in the parameter space of $\xi_0$, 
we are able to discuss which one is realized as a vacuum 
at the specified values of parameters $\xi$. 
The phase transition may also be argued on this basis 
in such a way that the state corresponding to the vacuum 
changes under the variation of the parameters $\xi$. 

It should be noted that 
although the free energy is a generating function of 
all observables which are obtained as derivatives 
with respect to corresponding parameters $\xi$, 
analyses of the improved perturbation theory should be 
applied independently to each observable of interest. 
This is because the prescription of improvement does not
commute with differentiation with respect to the parameters $\xi$  
and thus the formation of plateau may occur in a different 
manner by observables especially at lower orders of perturbation.

%%%%%%%%%%%%%%%%%%%%%%%%%%%%%%%%%%%%%%%%%%%%%%%%%%%%%%%%%%%%%%%%%
\section{Application to the massive Yang-Mills Chern-Simons matrix model}
%%%%%%%%%%%%%%%%%%%%%%%%%%%%%%%%%%%%%%%%%%%%%%%%%%%%%%%%%%%%%%%%%
\label{sec:mymcsmm}

In this section we discuss how to apply the improved perturbation 
theory to the massive YMCS matrix model. 
In order to determine the phase structure, we concentrate 
on the free energy $F(\alpha,m^2)$ given in Eq.~(\ref{free energy}) 
and the expectation value of the Chern-Simons term 
\begin{equation}
	A(\alpha,m^2)
	\equiv
	\frac{\partial F(\alpha,m^2)}{\partial \alpha}
	=
	\left\langle\frac{1}{N}\tr\left(
	\frac{1}{3}
	i\epsilon_{\mu\nu\lambda}A_{\mu}A_{\nu}A_{\lambda}\right)
	\right\rangle,
\end{equation}
as functions of two parameters $\alpha$, $m$ 
appearing in Eq.~(\ref{eq:action}). The reason we consider $A(\alpha,m^2)$ 
is that it would be an order parameter of our model. 
Namely, discussion in Section~\ref{sec:model} suggests that 
both $F(\alpha,m^2)$ and $A(\alpha,m^2)$ would be 
of $\mathcal{O}(1)$ or even smaller in the Yang-Mills phase, 
while they would become of $\mathcal{O}(1)$, and of $\mathcal{O}(N^2)$ 
in the 5-brane phase, and the fuzzy sphere phase respectively, 
corresponding to the fuzzy sphere type configuration 
realized in each phase.  

According to the general prescription given in the previous section, 
what we should do is first to introduce a coupling 
constant $\lambda$, which is set to be 1 in the end, 
into the original action (\ref{eq:action}) in order
to define a perturbative series as 
\begin{equation}
	S = 
	N\tr\left(-\frac{
        \lambda
	}{4}[A_{\mu},A_{\nu}]^2
	+\frac{i}{3}
        \sqrt{\lambda}
        \alpha
	\epsilon_{\mu\nu\lambda}A_{\mu}A_{\nu}A_{\lambda}
	+\frac{1}{2}
        m^2
	A_{\mu}^2\right),
\label{eq:YMCSaction}
\end{equation}
then to calculate $F(\alpha,m^2)$ or $A(\alpha,m^2)$ 
up to a certain order of $\lambda$. 
Since we are interested in the large-$N$ limit of our model, we fix 
the 't Hooft coupling $\lambda$ 
and consider only the planar diagrams.\footnote{%
In other words, we take the $N\to\infty$ limit in such a way 
that the coupling constants of four-point vertex and Chern-Simons term 
scale like Eq.~(\ref{eq:YMCSaction}) which is the same as 
in Refs.~\cite{Azuma:2005bj,Azuma:2004zq}.}
Next we make the replacement 
\begin{align}
	\lambda & \longrightarrow g\lambda, \nonumber \\ 
	\alpha  & \longrightarrow \alpha_0+g(\alpha-\alpha_0), \nonumber \\
	m^2     & \longrightarrow m_0^2+g(m^2-m_0^2)
\end{align}
in $F(\alpha,m^2)$ and $A(\alpha,m^2)$
to construct the improved perturbative series 
$F^\text{imp}(\alpha,m^2;\alpha_0,m_0^2)$ and 
$A^\text{imp}(\alpha,m^2;\alpha_0,m_0^2)$ for them. 
By searching for the plateau with respect to 
both $\alpha_0$ and $m_0^2$ 
introduced artificially, we can determine 
the true values of $F(\alpha,m^2)$ and $A(\alpha,m^2)$ for each 
$\alpha$ and $m^2$. 
 
In order to obtain the free energy, we have to count and calculate 
all diagrams up to an order we hope. 
In fact, there are many diagrams 
which contribute to the free energy even in lower orders. 
In order to avoid this difficulty, we follow the prescription 
given in 
Ref.~\cite{Kawai:2002jk}. 
Namely, the ordinary free energy is given 
by the Legendre transformation of the 2PI free energy 
given by the sum of diagrams in which there are no self-energy part 
and hence all propagators are the exact propagators.\footnote{%
The relation between the 2PI free energy and the 
Schwinger-Dyson equation is clarified in 
Ref.~\cite{Kawai:2002jk}.} 
In general, the number of diagrams which contribute 
to the 2PI free energy is drastically reduced 
compared to that of the ordinary free energy. 
This fact enables us to calculate the free energy 
in higher orders. In fact, we have performed it to the fifth order 
with respect to $\lambda$. 

For the purpose of calculating the 2PI free energy, 
we assume the form of the exact propagator as 
\begin{equation}
	\left\langle A_{\mu}{}^i{}_j A_{\nu}{}^k{}_l\right\rangle
	= 
	C\delta_{\mu\nu}{\delta^i}_l{\delta_j}^k.
\label{exact propagator}
\end{equation}
Then $C$ is the conjugate variable to $m^2$ 
in the sense of the Legendre transformation:
\begin{equation}
	\frac{\partial F}{\partial m^2}=\frac{3}{2}C,
	\qquad
	\frac{\partial G}{\partial C}=-\frac{3}{2}m^2,
\end{equation}
where $G$ is the 2PI free energy which is a function 
of $\alpha$ and $C$: $G=G(\alpha,C)$, from which we can obtain 
the free energy $F=F(\alpha,m^2)$ by the Legendre transformation 
with respect to $C$. 

Here we make a remark on the assumption (\ref{exact propagator}). 
If we allow the most generic form of the exact propagator 
\begin{equation}
	\left\langle A_{\mu}{}^i{}_j A_{\nu}{}^k{}_l\right\rangle
	= 
	C_{\mu\nu}{}^i{}_j{}^k{}_l,
\end{equation}
it has $6N^4$ parameters which become infinite in the large-$N$ limit. 
Correspondingly, we have infinite number of parameters 
conjugate to them in the free energy. This is equivalent to 
introducing the most generic quadratic term in the action 
as a mean field. Evidently this seems intractable. 
Therefore we follow the same approach as in 
Refs.~\cite{Nishimura:2001sx,Kawai:2002jk}. 
Namely, 
we impose appropriate symmetries on the exact propagator 
and assume its most generic form that is invariant under them. 
Then the number of the parameters in the exact propagator 
can be reduced to tractable one, and the free energy is written 
as a function of parameters allowed by the symmetries 
we have assumed. 
Thus we can find the most dominant configuration that respects them. 
In the present case let us assume the $\SO(3)$ Lorentz symmetry 
and the $\underbrace{\U(n)\times \U(n)\times \cdots \U(n)}_m\times S_m$ 
symmetry with $N=nm$. 
The latter symmetry reflects a ``clustering'' configuration 
\cite{Iso:1999xs}
in which there are $m$ clusters composed of $n$ eigenvalues 
and hence the $\U(n)$ gauge symmetry is expected to be generated. 
$S_m$ is the permutation of these $m$ clusters 
corresponding to the diffeomorphism on the fuzzy sphere. 
More precisely, the action of an element of $S_m$ on $A_{\mu}$ 
is defined by 
\begin{equation}
	{A_{\mu}}_{(ii')(jj')}
	\rightarrow 
	{A_{\mu}}_{(i\sigma(i'))(j\sigma(j'))}
	\qquad
	\text{for}\ \sigma\in S_m, 
\end{equation}
where $(ii')$ denotes the index $i$ of the $i'$-th $\U(n)$ group 
in $\underbrace{\U(n)\times \U(n)\times \cdots \U(n)}_m$. 
Then it is easy to see that the exact propagator 
is restricted to the form 
\begin{equation}
	C_{\mu\nu}{}^{(ii')} {}_{(jj')} {}^{(kk')} {}_{(ll')}
	=
	(C\delta_{\mu\nu}+D\delta_{\mu\nu}\delta_{i'j'})
	{\delta^i}_l{\delta_j}^k {\delta^{i'}}_{l'}{\delta_{j'}}^{k'},
	\qquad
	\text{not summed on}\ i',j'.
\label{general exact propagator}
\end{equation}
From this expression we see that contributions of 
diagrams containing the second term are in general suppressed by $1/m$. 
Therefore as long as $m$ is $\mathcal{O}(N)$, 
we do not have to take them into account in the large-$N$ limit 
and the exact propagator can be given by Eq.~(\ref{exact propagator}). 
In fact, this is the case in the fuzzy sphere phase 
in which there exists a single fuzzy sphere and hence $n=1$, $m=N$. 
On the other hand, the propagator should take the form 
in Eq.~(\ref{exact propagator}) also in the Yang-Mills phase, 
because the original $\U(N)$ gauge symmetry remains intact. 
However, in the 5-brane phase it is expected that 
there exist $\mathcal{O}(N)$ coincident small fuzzy spheres, 
which corresponds to the case 
$n\sim \mathcal{O}(N)$, $m\sim \mathcal{O}(1)$. 
Therefore, from the viewpoint in 
Refs.~\cite{Kawai:2002jk,Nishimura:2001sx} 
we have to include the parameter $D$ in Eq.~(\ref{general exact propagator}) 
in calculating the 2PI free energy, then the free energy would become 
a function of 3 parameters including one conjugate to $D$ 
other than $\alpha$ and $m$. 
In order to avoid such a situation, 
we simply assume that the contribution of $D$ 
in Eq.~(\ref{general exact propagator}) would be irrelevant and that 
even in the 5-brane phase the exact propagator would take the form 
in Eq.~(\ref{exact propagator}). 
Then we check consistency that 
the free energy $F(\alpha,m^2)$ and the expectation value of 
the Chern-Simons term $A(\alpha,m^2)$ derived from $G(\alpha, C)$ 
show behavior expected from the configuration in the 5-brane phase. 
Another reason we anticipate (\ref{exact propagator}) even in the 
5-brane phase is that 
from the standpoint of the noncommutative field theory, 
the original gauge symmetry is not broken, but is reinterpreted 
as the noncommutative gauge symmetry as shown in 
Refs.~\cite{Aoki:1999vr,Iso:2001mg}. 
In this sense, it is reasonable 
to assume (\ref{exact propagator}) in the Yang-Mills, 5-brane, 
and fuzzy sphere phases. 

Moreover, it is worth noticing that the improved perturbation theory 
with the ansatz in (\ref{exact propagator}) can investigate 
various vacua which are not restricted to three classical vacua 
as above as long as they satisfy the ansatz.\footnote
{However, the ansatz in (\ref{exact propagator}) does not take 
account of configurations considered in Ref.~\cite{Azuma:2005bj} 
where fuzzy spheres of various sizes appear.}

%================================================================
\section{Results}
%================================================================
\label{sec:result}

We calculate a perturbative series of 
the action (\ref{eq:YMCSaction}) up to the fifth order. 
From this series we obtain the improved series 
according to the prescription 
presented in Section~\ref{sec:improve}. 
We identify plateaux as accumulations of extrema of observables 
with respect to artificial parameters. 
In the case of a model which 
exhibits phase transition, distribution of 
extrema and the value of observables in a plateau region change 
drastically around a phase transition point \cite{Aoyama:2005nd}. 
Thus we examine the distributions of extrema 
of the improved series 
$F^\text{imp}(\alpha,m^2;\alpha_0,m_0^2)$ and 
$A^\text{imp}(\alpha,m^2;\alpha_0,m_0^2)$ with respect to 
$\alpha_0$ and $m_0^2$ 
for each $\alpha$ and $m^2$. 
In this paper we sweep 
$\alpha$ and $m^2$ from 0.0 to 10.0. 
This is because when these parameters are beyond 10.0 
we do not observe a drastic change in the distributions of 
extrema and the value of observables around the extrema. 

In the following, we begin with describing detailed examinations at 
four characteristic points in the parameter space $(\alpha,m^2)$.

\begin{itemize}
% - - - - - - - - - - - - - - - - - - - - - - - - - - - - - - - -
\item{$(\alpha,m^2)=(0.0,0.0)$}

\begin{figure}
\begin{center}
\subfigure[][]{%
	\includegraphics[scale=1.0]{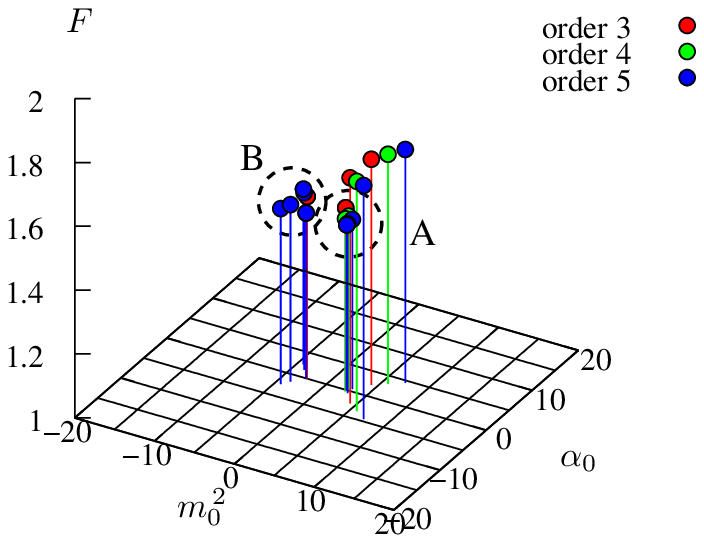}
	\label{fig:free0_0_3d}
}
\hskip 1em
\subfigure[][]{%
	\includegraphics[scale=.8]{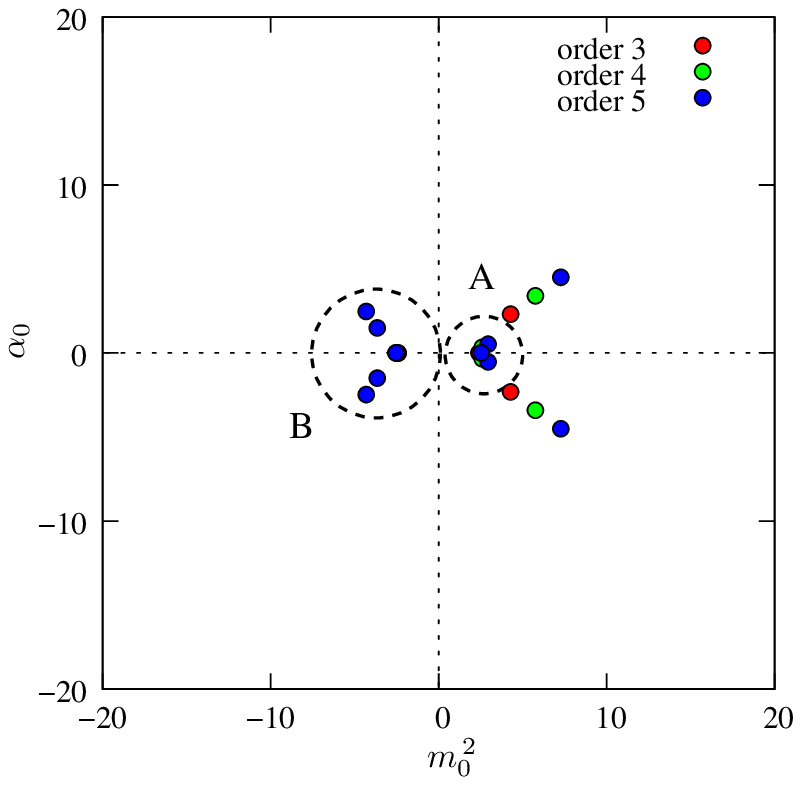}
	\label{fig:free0_0_2d}
}
\end{center}
\caption{%
Distribution of extrema of $F^\text{imp}(\alpha,m^2;\alpha_0,m_0^2)$ 
with respect to $m_0^2$ and $\alpha_0$ at $\alpha=0.0$ and $m^2=0.0$. 
Vertical lines indicate the value of $F^\text{imp}$ at the extrema 
\subref{fig:free0_0_3d}, 
and its 2-d plot in $(m_0^2,\alpha_0)$-plane \subref{fig:free0_0_2d}.  
}
\label{fig:free0_0}
\end{figure}

This corresponds to the pure Yang-Mills matrix model only with 
the four-point interaction terms. 

The distribution of extrema of $F^\text{imp}(\alpha_0,m_0^2)$  
with respect to $m_0^2$, $\alpha_0$ 
and its value at each extremum are shown in Fig.~\ref{fig:free0_0}. 
We can find two accumulations of extrema 
as shown by the dashed circles A, B. 
These two accumulations also appear for $A^\text{imp}(\alpha_0,m_0^2)$ 
and the improved series for the expectation value 
of the second moment of eigenvalues 
\begin{equation}
	C(\alpha,m^2)
	\equiv
	\frac{1}{3}\sum_\mu
	\left\langle
		\frac{1}{N}\tr\left( A_{\mu}^2\right) 
	\right\rangle. 
\label{eq:c-def}
\end{equation}
Except for these accumulations, there are gentle plateaux 
in the regions where the value of $|m_0^{\,2}|$ is very large. 
However, we do not take into account these flat regions as plateaux 
because these are nothing but asymptotic behaviors. 
These asymptotic behaviors always appear and we neglect them hereafter. 
By comparing the values of the improved free energy, 
we regard the plateau of region A as the true vacuum. 
The concrete values of the extrema 
and those of $F^\text{imp}(\alpha_0,m_0^2)$ 
there are listed in Table \ref{table:free0_0}, and 
those for $A^\text{imp}(\alpha_0,m_0^2)$ 
are listed in Table \ref{table:a0_0}. 
Because $A^\text{imp}$ is zero (Table \ref{table:a0_0}), 
this vacuum can be recognized as the Yang-Mills vacuum. 

Thus this parameter region corresponds to the Yang-Mills phase.

% - - - - - - - - - - - - - - - - - - - - - - - - - - - - - - - -
\item{$(\alpha,m^2)=(10.0,0.0)$}

\begin{figure}
\begin{center}
\subfigure[][]{%
	\includegraphics[scale=1.0]{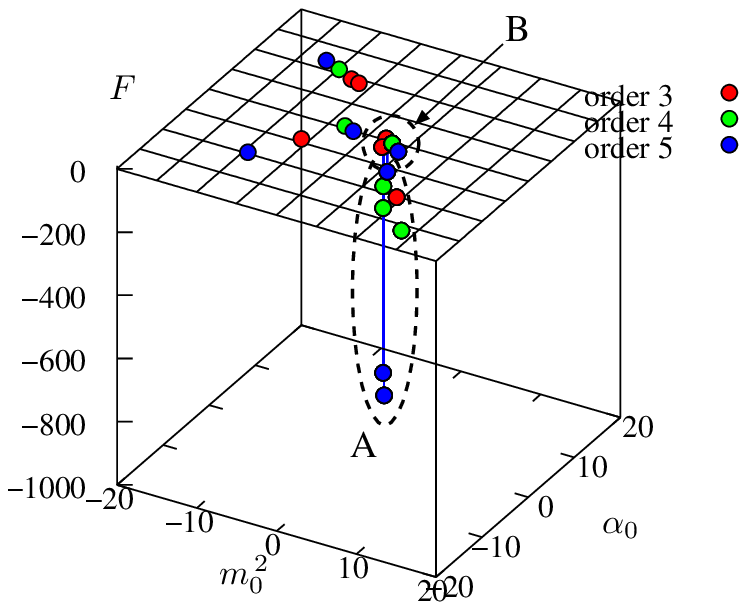}
	\label{fig:free0_10_3d}
}
\hskip 1em
\subfigure[][]{%
	\includegraphics[scale=0.8]{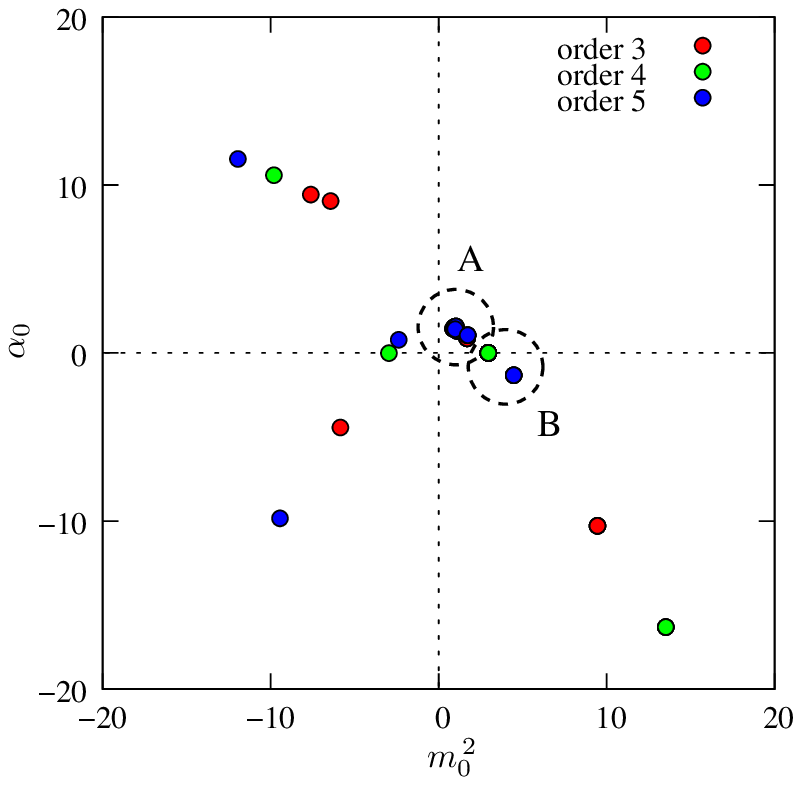}
	\label{fig:free0_10_2d}
}
\end{center}
\caption{%
Distribution of extrema of $F^\text{imp}(10.0,0.0;\alpha_0,m_0^2)$ 
with respect to $m_0^2$ and $\alpha_0$. 
Vertical lines indicate the value of $F^\text{imp}$ at the extrema 
\subref{fig:free0_10_3d}, 
and its 2-d plot in $(m_0^2,\alpha_0)$-plane \subref{fig:free0_10_2d}. 
}
\label{fig:free0_10}
\end{figure}

The distribution of extrema of $F^\text{imp}(\alpha_0,m_0^2)$ 
with respect to $m_0$, $\alpha_0$ 
and its value at each extremum are shown 
in Fig.~\ref{fig:free0_10}. 
From the Fig.~\ref{fig:free0_10_2d} we find an accumulation 
of extrema. We point out that this accumulation is divided 
into two sets A and B by examining $F^\text{imp}$. 
As for the region A, the values of $F^\text{imp}$ at extrema 
go to large negative values as the order increases. 
On the other hand the values of $F^\text{imp}$ at extrema 
in the region B remain $\mathcal{O}(1)$. 
They also appear for $A^\text{imp}$. 
The concrete values of the extrema 
and those of $F^\text{imp}(\alpha_0,m_0^2)$ 
there are listed in Table \ref{table:free0_10}, and 
those for $A^\text{imp}(\alpha_0,m_0^2)$ 
are listed in Table \ref{table:a0_10}. 

One might think that the region A cannot be regarded as 
a plateau because the values of $F^\text{imp}$ are not stable 
against the increase of the order 
(Fig.~\ref{fig:free0_10_3d}). 
These values are all beyond $\mathcal{O}(1)$ and 
in particular, the values of $F^\text{imp}$ and $A^\text{imp}$ 
at the fifth order reach to about $\mathcal{O}(10^3)$ 
and $\mathcal{O}(10^4)$, respectively,  
as shown in Table~\ref{table:free0_10} and Table~\ref{table:a0_10}. 
In general the improved perturbation 
theory is an approximation by 
polynomials of finite order and therefore 
the value of $\mathcal{O}(N^2)$ which goes 
infinite as $N \to \infty$ is expected to be approximated 
as large amount of value. As we go to higher orders, 
the approximation would become better 
and would provide a larger value in such a case. 
In this sense, instability of an approximated value 
against an increase of order is not always problematic. 
Thus we identify the region A as a plateau. 
By comparing the values of $F^\text{imp}$ in the region A and B 
we regard the plateau of the region A as the true vacuum. 

Because the value of  $A^\text{imp}$ in this 
plateau is $\mathcal{O}(10^4)$ (Table \ref{table:a0_10}), 
this vacuum can be recognized as the fuzzy sphere vacuum and this 
parameter region corresponds to the fuzzy sphere phase.

% - - - - - - - - - - - - - - - - - - - - - - - - - - - - - - - -
\item{$(\alpha,m^2)=(4.0,1.5)$}

\begin{figure}
\begin{center}
\subfigure[][]{%
	\includegraphics[scale=1.0]{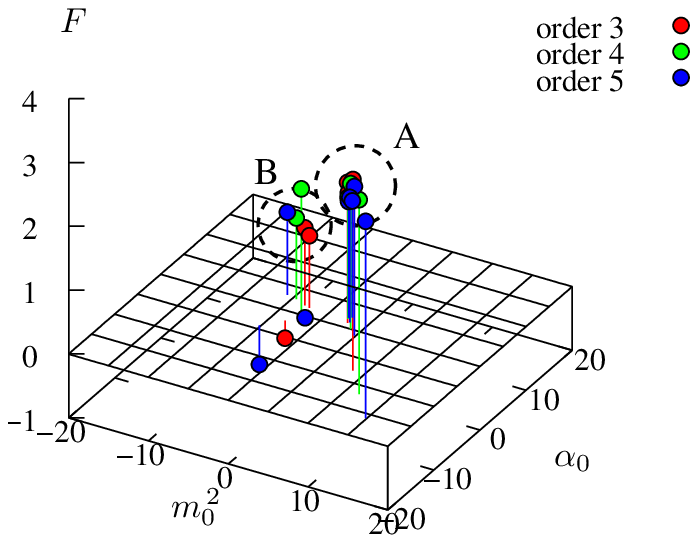}
	\label{fig:free2_2_3d}
}
\hskip 1em
\subfigure[][]{%
	\includegraphics[scale=0.8]{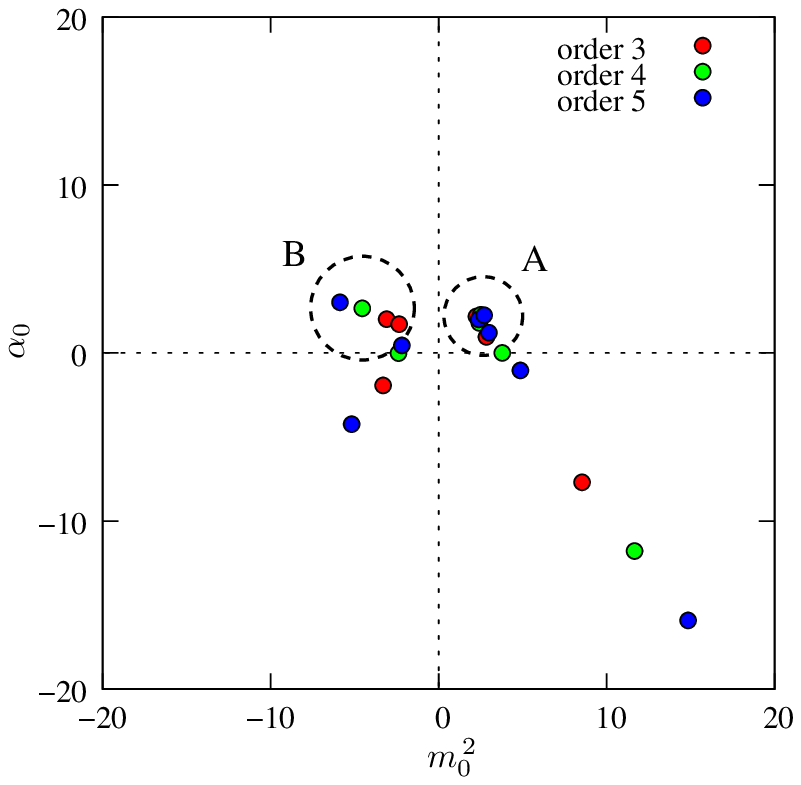}
	\label{fig:free2_2_2d}
}
\end{center}
\caption{%
Distribution of extrema of $F^\text{imp}(4.0,1.5;\alpha_0,m_0^2)$ 
with respect to $m_0^2$ and $\alpha_0$. 
Vertical lines indicate the value of $F^\text{imp}$ at the extrema 
\subref{fig:free2_2_3d}, 
and its 2-d plot in $(m_0^2,\alpha_0)$-plane \subref{fig:free2_2_2d}. 
}
\label{fig:free2_2}
\end{figure}
\begin{figure}
\begin{center}
\subfigure[][]{%
	\includegraphics[scale=1.0]{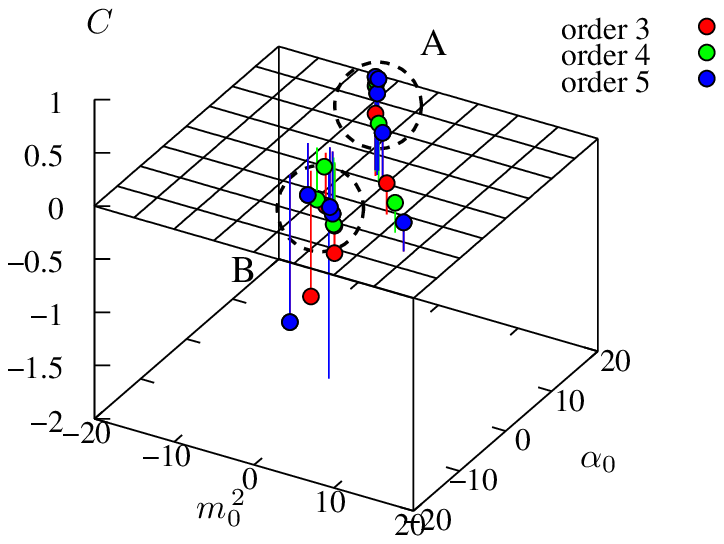}
	\label{fig:c2_2_3d}
}
\hskip 1em
\subfigure[][]{%
	\includegraphics[scale=0.8]{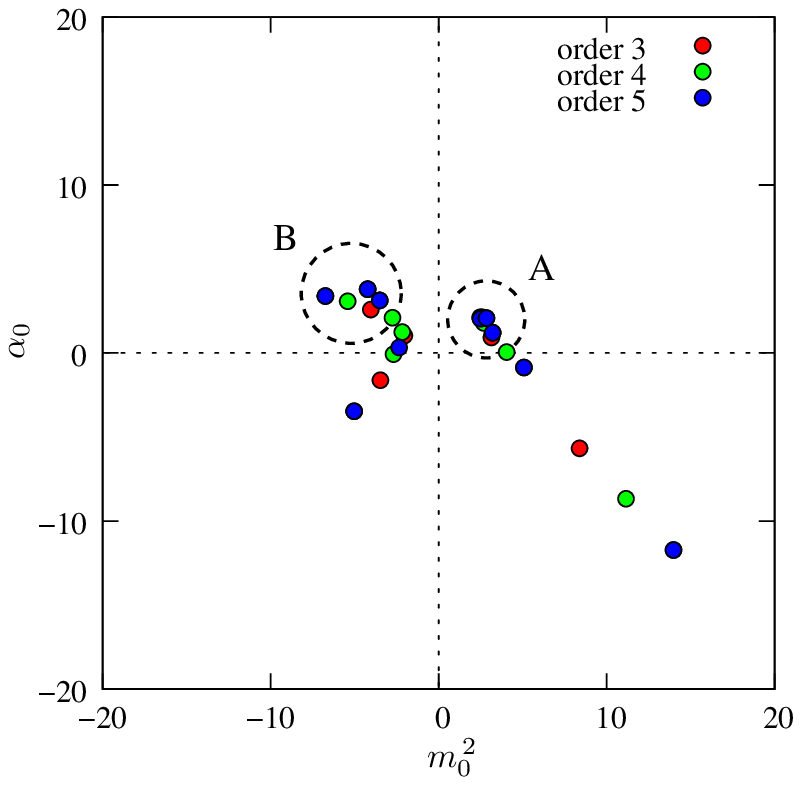}
	\label{fig:c2_2_2d}
}
\end{center}
\caption{%
Distribution of extrema of $C^\text{imp}(4.0,1.5;\alpha_0,m_0^2)$ 
with respect to $m_0^2$ and $\alpha_0$. 
Vertical lines indicate the value of $C^\text{imp}$ at the extrema 
\subref{fig:c2_2_3d}, 
and its 2-d plot in $(m_0^2,\alpha_0)$-plane \subref{fig:c2_2_2d}. 
}
\label{fig:c2_2}
\end{figure}

The distribution of extrema of $F^\text{imp}(\alpha_0,m_0^2)$ 
with respect to $m_0$, $\alpha_0$ 
and its value at each extremum are shown 
in Fig.~\ref{fig:free2_2}. 
We can find two accumulations of extrema. 
They also appear for $A^\text{imp}$. 
By comparing the values of $F^\text{imp}$, 
the region B is expected as the true vacuum. 
However the values of $C^\text{imp}$ at extrema in this plateau 
become negative (Fig.~\ref{fig:c2_2}). 
Thus we determine this plateau is unphysical. 
Then the plateau in the region A is recognized as the true vacuum 
which has negative $\mathcal{O}(1)$ values of $F^\text{imp}$ 
and $A^\text{imp}$. 
The concrete values of the extrema 
and those of $F^\text{imp}(\alpha_0,m_0^2)$, 
$C^\text{imp}(\alpha_0,m_0^2)$, and 
$A^\text{imp}(\alpha_0,m_0^2)$ are listed 
in Table \ref{table:free2_2}, \ref{table:c2_2}, and 
\ref{table:a2_2}, respectively. 

Because the value of $A^\text{imp}$ in the region A is negative 
$\mathcal{O}(1)$ value (Table \ref{table:a2_2}) 
this parameter region corresponds to the 5-brane phase.

% - - - - - - - - - - - - - - - - - - - - - - - - - - - - - - - -
\item{$(\alpha,m^2)=(2.0,10.0)$}

\begin{figure}
\begin{center}
\subfigure[][]{%
	\includegraphics[scale=1.0]{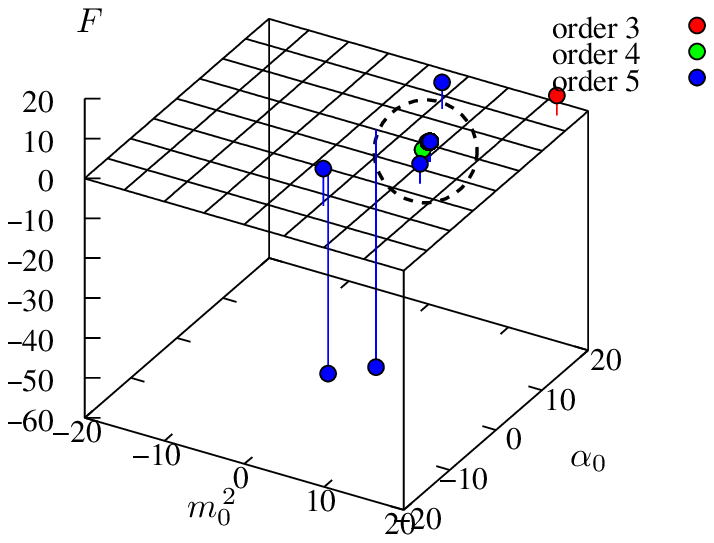}
	\label{fig:free10_2_3d}
}
\hskip 1em
\subfigure[][]{%
	\includegraphics[scale=0.8]{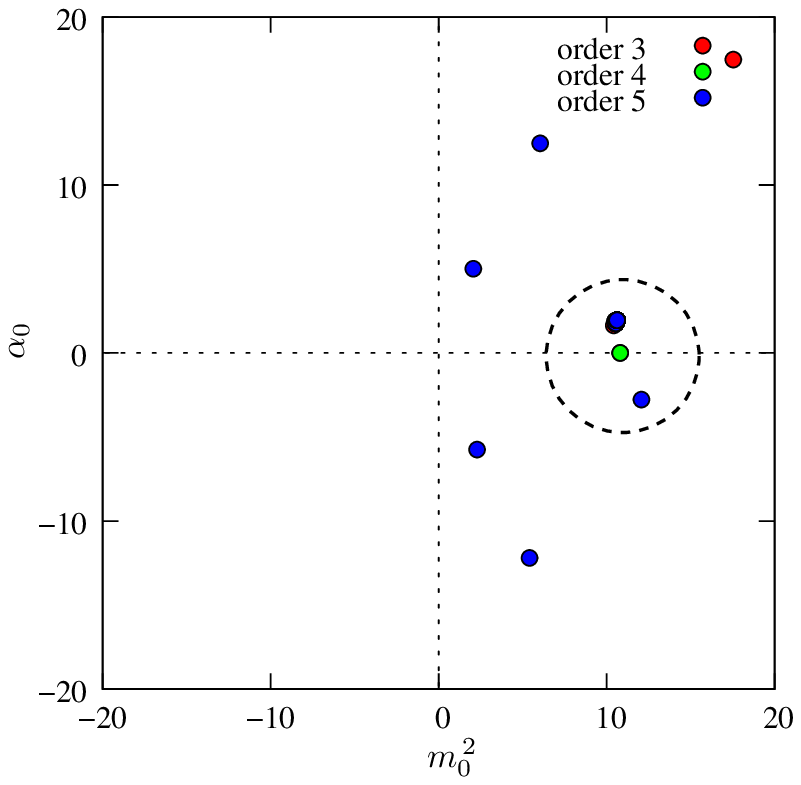}
	\label{fig:free10_2_2d}
}
\end{center}
\caption{%
Distribution of extrema of $F^\text{imp}(2.0,10.0;\alpha_0,m_0^2)$ 
with respect to $m_0^2$ and $\alpha_0$. 
Vertical lines indicate the value of $F^\text{imp}$ at the extrema 
\subref{fig:free10_2_3d}, 
and its 2-d plot in $(m_0^2,\alpha_0)$-plane \subref{fig:free10_2_2d}. 
}
\label{fig:free10_2}
\end{figure}

The distribution of extrema of $F^\text{imp}(\alpha_0,m_0^2)$ 
with respect to $m_0$, $\alpha_0$ 
and its value at each extremum are shown 
in Fig.~\ref{fig:free10_2}. 
We can find an accumulation of extrema. 
The concrete values of the extrema 
and those of $F^\text{imp}(\alpha_0,m_0^2)$ 
there are listed in Table \ref{table:free10_2}, and 
those for $A^\text{imp}(\alpha_0,m_0^2)$ 
are listed in Table \ref{table:a10_2}. 

Because the value of  $A^\text{imp}$ in this 
plateau is around zero (Table \ref{table:a10_2}), 
this vacuum corresponds to the Yang-Mills vacuum. 
Thus this parameter region corresponds to the Yang-Mills phase. 

% - - - - - - - - - - - - - - - - - - - - - - - - - - - - - - - -
\end{itemize}

\begin{figure}
\begin{center}
    \includegraphics[scale=.75]{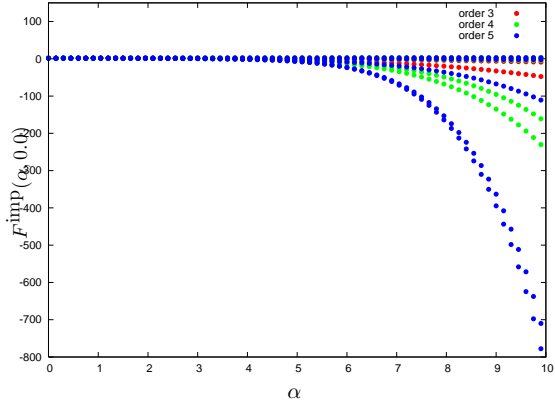}
\end{center}
\caption{%
Values of $F^\text{imp}(\alpha,0.0;\alpha_0,m_0^2)$ 
at extrema as functions of $\alpha$.}
\label{fig:yoko_F}
\end{figure}

\begin{figure}
\begin{center}
\subfigure[][]{%
	\includegraphics[scale=.75]{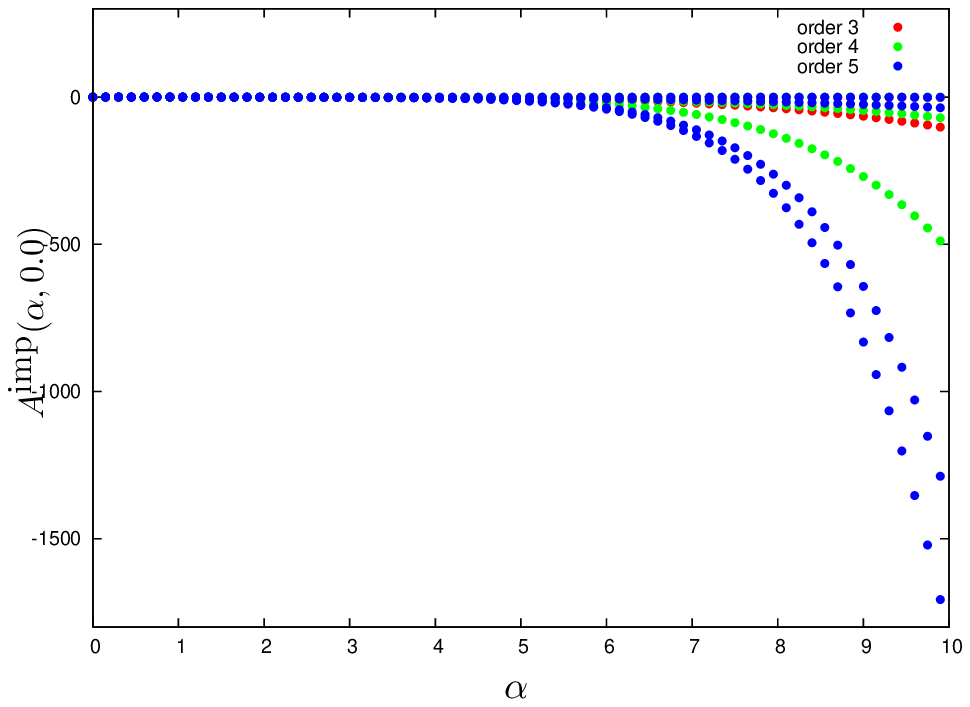}
	\label{fig:yokonormal}
}
\hskip 1em
\subfigure[][]{%
	\includegraphics[scale=.75]{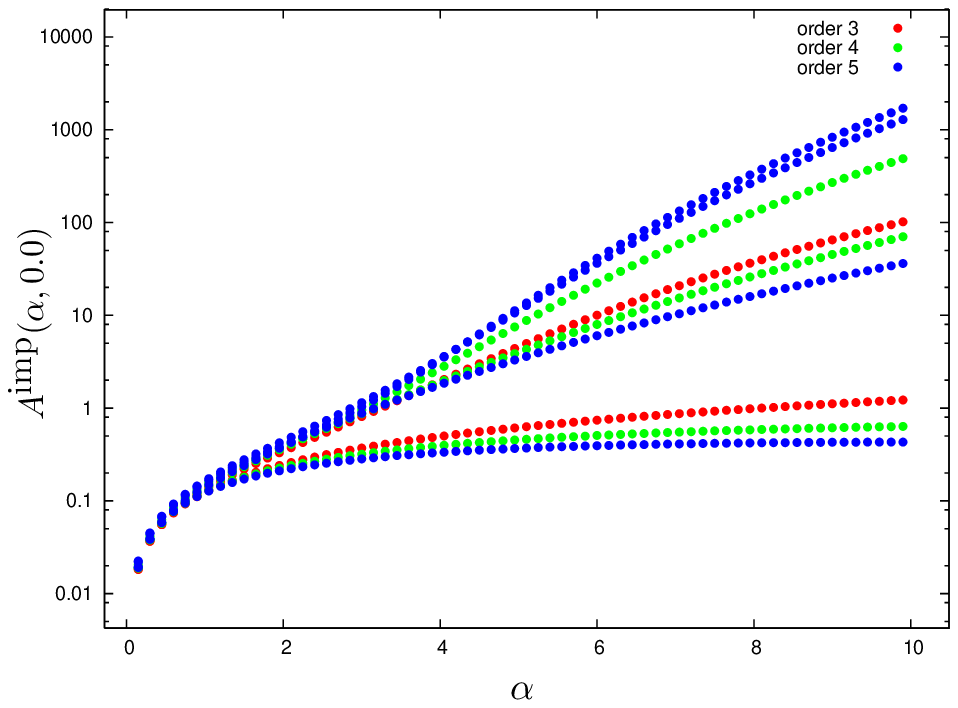}
	\label{fig:yokolog}
}
\end{center}
\caption{%
Values of $A^\text{imp}(\alpha,0.0;\alpha_0,m_0^2)$ 
at extrema \subref{fig:yokonormal} 
and its absolute value in logarithmic scale \subref{fig:yokolog} 
as functions of $\alpha$.}
\label{fig:yoko}
\end{figure}

Now we proceed to discussion in the case of generic $\alpha$ and $m^2$. 
As an illustration 
we plot the values of $F^\text{imp}$ and $A^\text{imp}$ 
at their extrema 
with positive\footnote{%
More precisely, we chose the region 
$m_0^2=0\sim 20$, $\alpha_0=-20\sim 20$. This is because 
there is no accumulation except for foregoing asymptotic behavior 
in the region of $|\alpha_0|,|m_0^2| \ge 20$.
}
$m_0^2$ as functions of $\alpha$ when $m^2$ is zero 
(Fig.~\ref{fig:yoko_F} and Fig.~\ref{fig:yoko}). 
From these figures we can divide the set of extrema 
into two branches: 
one branch consists of extrema for which 
the values of $F^\text{imp}$ and $A^\text{imp}$ 
remain of $\mathcal{O}(1)$ against the increase of $\alpha$, 
while the other branch is composed of extrema 
for which the values of $F^\text{imp}$ and $A^\text{imp}$ decrease 
far more than of $\mathcal{O}(10)$ as $\alpha$ increases. 
This property can be seen clearly in Fig.~\ref{fig:yokolog}. 
We call the former as branch B and the latter as branch A. 
In fact each branch corresponds to an accumulation of extrema 
in $(\alpha_0,m_0^2)$-plane. 
The branch A, B corresponds to the plateau of 
the region A, B respectively at the above mentioned case 
$(\alpha,m^2)=(10.0,0.0)$. 
These two branches join together for small $\alpha$, 
and thus we find only one accumulation at $(\alpha,m^2)=(0.0,0.0)$. 

By comparing the values of $F^\text{imp}$ (Fig.~\ref{fig:yoko_F}), 
we identify the plateau corresponding to the branch A 
as the true vacuum as in the case of $(\alpha,m^2)=(10.0,0.0)$. 
We take the expectation value of the Chern-Simons term 
as an order parameter. 
By examining the value of $A^\text{imp}$ in this branch 
we can determine what phase appears at each $(\alpha,m^2)$. 
If it takes a value of $\mathcal{O}(N^2)$, of $\mathcal{O}(1)$, 
and around zero, it corresponds to the fuzzy sphere phase, 
5-brane phase, and Yang-Mills phase, respectively. 
Thus from Fig.~\ref{fig:yokolog} 
we conclude that 
there is the Yang-Mills phase in the region $0 < \alpha < 2$, 
the 5-brane phase in the region $2 < \alpha < 4$ 
and the fuzzy sphere phase in the region $\alpha > 4$. 
We also note that if the set of extrema is divided into branches, 
it indicates that the behavior of the improved series 
and the distribution of extrema change drastically there. 
Then this would imply a phase transition there 
as mentioned in the beginning of this section. 
In fact, Fig.~\ref{fig:yokolog} shows this is the case with our model, 
namely around the phase transition points we can observe 
the branching of the extrema. 

\begin{figure}
\begin{center}
	\includegraphics[scale=.75]{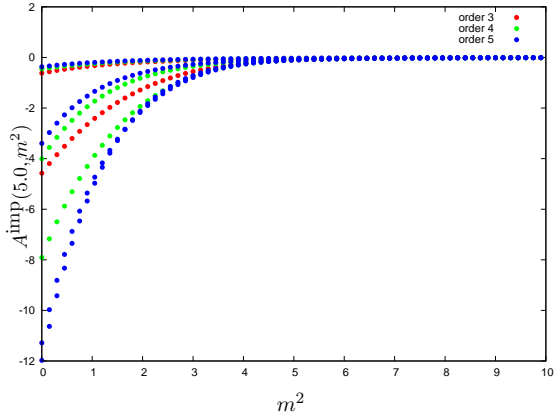}
\end{center}
\caption{%
Values of $A^\text{imp}(5.0,m^2;\alpha_0,m_0^2)$ at extrema 
as functions of $m^2$.}
\label{fig:tate}
\end{figure}

\begin{figure}
\begin{center}
	\includegraphics[scale=.8]{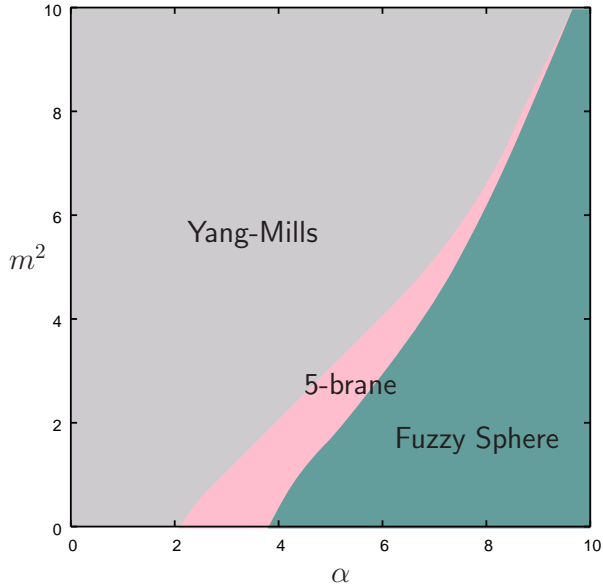}
\end{center}
\caption{%
A proposed phase diagram of the massive Yang-Mills Chern-Simons matrix model.}
\label{fig:phasediagram}
\end{figure}

In the same manner, we observe the behavior of 
the value at extrema of $A^\text{imp}$ 
along an arbitrary path 
in $(\alpha,m^2)$-plane. As an example, we show values 
of $A^\text{imp}(\alpha,m^2;\alpha_0,\alpha_0^2)$ 
at extrema as functions of $m^2$ when $\alpha=5.0$ in 
Fig.~\ref{fig:tate}.
From these investigations 
we can obtain a phase diagram of the massive YMCS matrix model 
in $(\alpha,m^2)$-plane (Fig~\ref{fig:phasediagram}). 

%================================================================
\section{Conclusions and Discussions}
%================================================================
\label{sec:discussion}

In this paper we applied the improved perturbation theory 
to the massive YMCS matrix model 
and calculated the free energy, and the expectation values 
of the second moment of eigenvalues and the Chern-Simons term 
up to the fifth order. 
Though the original perturbative series was obtained about 
the perturbative vacuum corresponding to the Yang-Mills vacuum, 
the improved perturbation theory reveals the information 
on the different vacua, 
namely the 5-brane vacuum and the fuzzy sphere vacuum. 

As a result, we obtained the phase diagram of this model. 
The results are very similar to the previous work based on 
the one-loop calculation around the fuzzy sphere type background. 
Since the improved perturbation theory is expected to 
include nonperturbative effects, 
we can expect that a gauge symmetry is generated 
as a subgroup of the original $\U(N)$ group 
even nonperturbatively 
in the 5-brane phase and the fuzzy sphere phase. 
In particular, dynamical generation of a non-Abelian gauge group 
occurs nonperturbatively in the 5-brane phase. 
However the precise positions of the critical lines 
in our phase diagram are somewhat ambiguous. 
For example, in our phase diagram 
there exists the 5-brane phase for $2<\alpha<4$ when $m^2=0$, 
which apparently contradicts with the Monte Carlo result 
in Ref.~\cite{Azuma:2004zq}. 
This is because our improved perturbative series is constructed 
from the standard perturbative series of lower order. 
It is obvious that 
in order to observe the phase transition clearly, 
a lower order analysis is not sufficient. 
As we proceed to higher orders, this ambiguity would decrease 
and we would find the phase transitions between the three phases 
more precisely. Then a precise comparison to the phase diagram 
proposed in Ref.~\cite{Azuma:2005bj} makes sense and we can 
discuss to what extent the perturbative analysis is valid 
in various parameter regions. 

Another result of this paper is a confirmation of validity 
of the improved perturbation theory. 
It is accomplished by comparing our result in Fig.~\ref{fig:yoko} 
with the precedent result of the Monte Carlo method \cite{Azuma:2004yg}. 
Although we can not see clearly the first-order phase transition 
between the Yang-Mills phase and the fuzzy sphere phase, 
our result indicates a similar parameter region 
($\alpha_{critical}\sim \mathcal{O}(1)$) 
as a critical point to that of the Monte Carlo method.

It is worthwhile to mention the accumulation which was 
regarded as an unphysical plateau when $(\alpha,m^2)=(4.0,1.5)$. 
This accumulation of the extrema shows 
the lower value of the free energy than that of the other plateau. 
Therefore it would be the true vacuum 
if it corresponds to a physical state. 
It is interesting that it has a positive expectation value 
of the Chern-Simons term which is an opposite sign 
compared to that in the usual 5-brane or fuzzy sphere vacuum. 
We considered this plateau unphysical 
because the expectation value of the second moment of eigenvalues 
is negative there. 
However if it becomes positive when 
we go to a higher order or relax the $\SO(3)$-symmetric assumption 
(\ref{exact propagator}), then we would have a nontrivial phase 
and the phase diagram would change. 
We leave it to a future work.

%================================================================
\begin{acknowledgments}
The authors are grateful to 
T.~Azuma, 
H.~Kawai, 
T.~Matsuo,  
and 
J.~Nishimura 
for valuable discussions. 
T.~K and Y.~S. are supported by 
the Special Postdoctoral Researchers Program at RIKEN. 
\end{acknowledgments}
%================================================================

%================================================================

%================================================================

\newpage

\appendix

%\section{Numerical values}

\renewcommand{\baselinestretch}{0.8}
\begin{table}
\begin{tabular}{cddd}
\hline\hline
\multicolumn{1}{c}{\makebox[5em]{order}} &
\multicolumn{1}{c}{\makebox[12em]{$\alpha_0$}} &
\multicolumn{1}{c}{\makebox[12em]{$m_0^{\,2}$}} &
\multicolumn{1}{c}{\makebox[12em]{$F^\text{imp}$}} \\
\hline
3
& -4.68729 \times 10^{9} & -8.30606 \times 10^{7} &  26.1027 \\
& 4.68729 \times 10^{9}  & -8.30606 \times 10^{7} &  26.1027 \\
& 0.0 & -2.40953 &   1.5702 \\
& 0.0 &  2.40953 &   1.5702 \\
& -2.31723 &  4.29037 &  1.70658 \\
&  2.31723 &  4.29037 &  1.70658 \\
\hline
4
& 0.341271 &  2.59243 &  1.54085 \\
& -0.341271 &  2.59243 &  1.54085 \\
&  3.40785 &  5.75606 &  1.71942 \\
& -3.40785 &  5.75606 &  1.71942 \\
\hline
5
& -2.46938 & -4.30854 &  1.54941 \\
&  2.46938 & -4.30854 &  1.54941 \\
& -1.49032 & -3.65313 &  1.55424 \\
&  1.49032 & -3.65313 &  1.55424 \\
& 0.0 & -2.54153 &  1.51756 \\
& 0.0 &  2.54153 &  1.51756 \\
& -0.524011 &   2.9491 &   1.5305 \\
& 0.524011 &   2.9491 &   1.5305 \\
& -4.51448 &  7.27529 &  1.73146 \\
&  4.51448 &  7.27529 &  1.73146 \\
\hline\hline
\end{tabular}
\caption{The numerical data of extrema for $F^\text{imp}(0.0,0.0;\alpha_0,m_0^2)$}
\label{table:free0_0}
\end{table}
\renewcommand{\baselinestretch}{1.5}

\renewcommand{\baselinestretch}{0.8}
\begin{table}
\begin{tabular}{cddd}
\hline\hline
\multicolumn{1}{c}{\makebox[5em]{order}} & 
\multicolumn{1}{c}{\makebox[12em]{$\alpha_0$}} & 
\multicolumn{1}{c}{\makebox[12em]{$m_0^{\,2}$}} & 
\multicolumn{1}{c}{\makebox[12em]{$A^\text{imp}$}} \\
\hline
3
& 0.0 & -3.08221 & 0.0 \\
& -0.138803 & -2.94881 & 0.000320748 \\
& 0.138803 & -2.94881 & -0.000320748 \\
& 0.0 & -2.82843 & 0.0 \\
& 0.0 &  2.82843 & 0.0 \\
& 0.0 &  3.08221 & 0.0 \\
\hline
4
&  5.45504 & -6.42054 & 0.116302 \\
& -5.45504 & -6.42054 & -0.116302 \\
& 0.0 &  -3.7272 & 0.0 \\
& -0.290656 &   -3.458 & 0.000734582 \\
& 0.290656 &   -3.458 & -0.000734582 \\
& 0.629447 & -3.26122 & 0.000832601 \\
& -0.629447 & -3.26122 & -0.000832601 \\
& 0.0 & -3.24237 & 0.0 \\
& 0.270866 & -2.89579 & 0.00388426 \\
& -0.270866 & -2.89579 & -0.00388426 \\
& 0.0 & -2.66364 & 0.0 \\
& 0.0 &  2.66364 & 0.0 \\
& 0.0 &  3.24237 & 0.0 \\
& 0.0 &   3.7272 & 0.0 \\
\hline
5
& 0.0 & -4.42101 & 0.0 \\
& 0.445076 & -4.03236 & -0.000992029 \\
& -0.445076 & -4.03236 & 0.000992029 \\
& 0.0 & -3.75696 & 0.0 \\
& 0.965698 & -3.72636 & 0.000171623 \\
& -0.965698 & -3.72636 & -0.000171623 \\
& -1.00261 & -3.69771 & -0.000168316 \\
&  1.00261 & -3.69771 & 0.000168316 \\
& 0.0787589 &  2.77319 & -0.000243519 \\
& -0.0787589 &  2.77319 & 0.000243519 \\
& 0.0 &  3.75696 & 0.0 \\
& 0.0 &  4.42101 & 0.0 \\
\hline\hline
\end{tabular}
\caption{The numerical data of extrema for $A^\text{imp}(0.0,0.0;\alpha_0,m_0^2)$}
\label{table:a0_0}
\end{table}
\renewcommand{\baselinestretch}{1.5}

\renewcommand{\baselinestretch}{0.8}
\begin{table}
\begin{tabular}{cddd}
\hline\hline
\multicolumn{1}{c}{\makebox[5em]{order}} & 
\multicolumn{1}{c}{\makebox[12em]{$\alpha_0$}} & 
\multicolumn{1}{c}{\makebox[12em]{$m_0^{\,2}$}} & 
\multicolumn{1}{c}{\makebox[12em]{$F^\text{imp}$}} \\
\hline
3
&  9.43649 & -7.60688 &  3.07945 \\
&  9.04356 & -6.42828 &  3.06715 \\
& -4.43165 & -5.84631 &  1.67693 \\
&  1.43896 & 0.862241 & -49.4706 \\
& 0.876274 &   1.6869 & -9.32175 \\
& -10.2822 &  9.45705 &  2.64061 \\
\hline
4
&  10.5828 & -9.79928 &  3.08573 \\
& -0.00542322 & -2.96341 &  7.73904 \\
&  1.53846 & 0.937571 & -243.211 \\
&  1.32581 &  1.08363 & -169.926 \\
& 0.00542358 &  2.95278 & -4.62466 \\
& -16.3116 &   13.527 &  2.75701 \\
\hline
5
&  11.5522 & -11.9498 &  3.08677 \\
& -9.84509 & -9.44094 &   1.5811 \\
& 0.791191 & -2.37935 & -14.4279 \\
&  1.42496 & 0.986792 & -762.155 \\
&  1.58502 &  1.01745 & -835.662 \\
&  1.06366 &  1.73127 &  -116.83 \\
& -1.32478 &  4.47375 &  -2.3464 \\
& -22.4794 &  17.6733 &  2.82902 \\
\hline\hline
\end{tabular}
\caption{The numerical data of extrema for $F^\text{imp}(10.0,0.0;\alpha_0,m_0^2)$}
\label{table:free0_10}
\end{table}
\renewcommand{\baselinestretch}{1.5}
\renewcommand{\baselinestretch}{0.8}
\begin{table}
\begin{tabular}{cddd}
\hline\hline
\multicolumn{1}{c}{\makebox[5em]{order}} & 
\multicolumn{1}{c}{\makebox[12em]{$\alpha_0$}} & 
\multicolumn{1}{c}{\makebox[12em]{$m_0^{\,2}$}} & 
\multicolumn{1}{c}{\makebox[12em]{$A^\text{imp}$}} \\
\hline
3
&  9.94959 & -7.34049 & 0.0901421 \\
& -0.00271556 & -2.95803 &  1.23634 \\
&  1.40096 & 0.995306 & -106.684 \\
& 0.00271758 &  2.95805 & -1.23634 \\
\hline
4
& -22.1864 & -15.5176 & -0.0340077 \\
&  11.4184 & -9.11834 & 0.0886916 \\
& 0.798254 &  -2.2225 & -8.76388 \\
&    1.501 & 0.998731 & -519.979 \\
&  1.12189 &  1.54166 & -73.7818 \\
& -1.65882 &  4.95124 & -0.637283 \\
\hline
5
&  12.4504 & -10.8361 & 0.0890299 \\
&  8.65517 & -6.44966 & 0.101811 \\
& -1.90908 & -4.94843 &  1.46692 \\
&   1.1106 & -2.05508 &  40.6892 \\
&  1.51445 &  1.01601 & -1385.51 \\
&  1.42806 &  1.18735 & -1840.99 \\
& 0.703556 &  2.33845 & -37.6504 \\
& -3.67883 &  7.04633 & -0.431045 \\
\hline\hline
\end{tabular}
\caption{The numerical data of extrema for $A^\text{imp}(10.0,0.0;\alpha_0,m_0^2)$}
\label{table:a0_10}
\end{table}
\renewcommand{\baselinestretch}{1.5}

\renewcommand{\baselinestretch}{0.8}
\begin{table}
\begin{tabular}{cddd}
\hline\hline
\multicolumn{1}{c}{\makebox[5em]{order}} & 
\multicolumn{1}{c}{\makebox[12em]{$\alpha_0$}} & 
\multicolumn{1}{c}{\makebox[12em]{$m_0^{\,2}$}} & 
\multicolumn{1}{c}{\makebox[12em]{$F^\text{imp}$}} \\
\hline
3
& 9.61196 \times 10^{10} & -6.49178 \times 10^{7} &  25.8232 \\
& -5.7603 \times 10^{9} & -4.47955 \times 10^{6} &  22.5561 \\
& -1.92635 &  -3.3072 & -0.277788 \\
&  2.01317 & -3.09678 &  1.21161 \\
&  1.71943 & -2.35192 &   1.1326 \\
&  2.17535 &  2.25552 &  1.94668 \\
& 0.966819 &  2.85037 &  2.20386 \\
& -7.70211 &  8.53538 &  2.99592 \\
& 1.82382 \times 10^{12} & 1.63535 \times 10^{8} &  8.64457 \\
\hline
4
&  2.66644 &  -4.5446 &  1.27361 \\
& -0.00980172 & -2.38145 &  1.97284 \\
&   1.7947 &  2.42488 &  1.90002 \\
&  2.27369 &  2.50761 &  1.86237 \\
& 0.0104127 &  3.78695 &  2.27765 \\
&  -11.785 &  11.6617 &  3.04714 \\
\hline
5
&  3.01798 & -5.86682 &  1.29304 \\
& -4.23407 & -5.18175 & -0.613976 \\
&  0.45086 & -2.18319 & -0.0677834 \\
&  2.00342 &  2.38336 &  1.82213 \\
&   2.2562 &   2.7082 &  1.82699 \\
&  1.20384 &  2.99017 &  1.95765 \\
& -1.03239 &  4.86228 &   2.3355 \\
& -15.9168 &  14.8553 &  3.08096 \\
\hline\hline
\end{tabular}
\caption{The numerical data of extrema for $F^\text{imp}(4.0,1.5;\alpha_0,m_0^2)$}
\label{table:free2_2}
\end{table}
\renewcommand{\baselinestretch}{1.5}
\renewcommand{\baselinestretch}{0.8}
\begin{table}
\begin{tabular}{cddd}
\hline\hline
\multicolumn{1}{c}{\makebox[5em]{order}} & 
\multicolumn{1}{c}{\makebox[12em]{$\alpha_0$}} & 
\multicolumn{1}{c}{\makebox[12em]{$m_0^{\,2}$}} & 
\multicolumn{1}{c}{\makebox[12em]{$C^\text{imp}$}} \\
\hline
3
&  3.39189 & -6.74048 & -0.48585 \\
& -3.47241 & -5.02842 & -1.38562 \\
&  3.80442 &  -4.2263 & -0.560315 \\
&  2.58809 & -4.03884 & -0.497245 \\
&   3.1278 & -3.50164 & -0.581923 \\
&   -1.621 & -3.47036 & -1.18014 \\
&   1.0354 & -2.03806 & -0.841414 \\
&  2.12207 &  2.47803 & 0.786263 \\
&  2.04249 &  2.50236 & 0.873765 \\
&  2.07634 &  2.85198 & 0.859968 \\
& 0.919071 &  3.14954 & 0.586119 \\
&  1.21074 &   3.2149 & 0.766484 \\
& -0.859116 &   5.0785 & 0.514263 \\
& -5.67532 &  8.38804 & 0.292278 \\
& -11.7333 &  13.9904 & 0.273057 \\
& 3.87597 \times 10^{12} & 4.55414 \times 10^{8} & 1.73042 \times 10^{-8} \\
& -4.05022 \times 10^{12} & 3.90586 \times 10^{9} & 0.0 \\
\hline
4
&  3.07677 & -5.41897 & -0.488433 \\
&   2.0976 & -2.75252 & -0.626686 \\
& -0.0734577 & -2.68203 & -0.00223366 \\
&  1.25144 & -2.16141 &  -0.5948 \\
&  2.12631 &  2.58726 & 0.829878 \\
&  1.79011 &  2.68894 & 0.811098 \\
& 0.0567346 &  4.05283 &  0.54443 \\
& -8.67938 &  11.1616 & 0.280595 \\
& -2.06462 \times 10^{11} & 5.2415 \times 10^{7} & 1.0194 \times 10^{-6} \\
& 2.78288 \times 10^{11} & 7.59018 \times 10^{7} & 2.4174 \times 10^{-7} \\
\hline
5
&  3.39189 & -6.74048 & -0.48585 \\
& -3.47241 & -5.02842 & -1.38562 \\
&  3.80442 &  -4.2263 & -0.560315 \\
&   3.1278 & -3.50164 & -0.581923 \\
& 0.330493 & -2.35159 & -2.49497 \\
&  2.04249 &  2.50236 & 0.873765 \\
&  2.07634 &  2.85198 & 0.859968 \\
&  1.21074 &   3.2149 & 0.766484 \\
& -0.859116 &   5.0785 & 0.514263 \\
& -11.7333 &  13.9904 & 0.273057 \\
& -1.2577 \times 10^{11} & 4.10542 \times 10^{7} & 2.22914 \times 10^{-6} \\
& 5.85754 \times 10^{11} & 1.06032 \times 10^{8} & 1.83291 \times 10^{-6} \\
\hline\hline
\end{tabular}
\caption{The numerical data of extrema for $C^\text{imp}(4.0,1.5;\alpha_0,m_0^2)$}
\label{table:c2_2}
\end{table}
\renewcommand{\baselinestretch}{1.5}
\renewcommand{\baselinestretch}{0.8}
\begin{table}
\begin{tabular}{cddd}
\hline\hline
\multicolumn{1}{c}{\makebox[5em]{order}} & 
\multicolumn{1}{c}{\makebox[12em]{$\alpha_0$}} & 
\multicolumn{1}{c}{\makebox[12em]{$m_0^{\,2}$}} & 
\multicolumn{1}{c}{\makebox[12em]{$A^\text{imp}$}} \\
\hline
3
&  2.34863 & -2.98275 & 0.344616 \\
& -0.00869776 & -2.30156 &  1.30642 \\
&  1.98521 &  2.40399 & -0.679792 \\
& 0.00529296 &  3.80166 & -0.198496 \\
\hline
4
& -8.43836 & -7.87598 & -0.122983 \\
&  3.58394 & -4.43987 & 0.312971 \\
& 0.553466 & -2.13643 & -0.541593 \\
&  2.11319 &  2.50874 & -0.812064 \\
&  1.33838 &  2.85121 & -0.576061 \\
& -1.28207 &  5.19139 & -0.156768 \\
\hline
5
&  4.11377 & -5.56741 & 0.310493 \\
&  -1.1206 & -3.34363 &   2.1189 \\
& 0.722978 & -2.04584 &  2.72295 \\
&  1.04265 & -2.01922 & 0.734287 \\
&  2.08655 &  2.54758 & -0.866595 \\
&  1.93161 &  2.68347 & -0.877446 \\
& 0.744671 &   3.4634 & -0.504538 \\
& -2.62651 &  6.65831 & -0.133145 \\
\hline\hline
\end{tabular}
\caption{The numerical data of extrema for $A^\text{imp}(4.0,1.5;\alpha_0,m_0^2)$}
\label{table:a2_2}
\end{table}
\renewcommand{\baselinestretch}{1.5}

\renewcommand{\baselinestretch}{0.8}
\begin{table}
\begin{tabular}{cddd}
\hline\hline
\multicolumn{1}{c}{\makebox[5em]{order}} & 
\multicolumn{1}{c}{\makebox[12em]{$\alpha_0$}} & 
\multicolumn{1}{c}{\makebox[12em]{$m_0^{\,2}$}} & 
\multicolumn{1}{c}{\makebox[12em]{$F^\text{imp}$}} \\
\hline
3
&  1.64718 &  10.4191 &  4.97903 \\
&  17.4705 &  17.5462 &  4.99209 \\
& -24.2305 &  20.3698 &  5.03619 \\
\hline
4
&  1.91975 &  10.5121 &  4.97903 \\
& 0.00745054 &  10.8081 &  4.97907 \\
& 0.0074495 &  10.8081 &  4.97907 \\
&  26.7536 &  22.5472 &  5.00089 \\
&   -36.51 &  26.4821 &  5.05249 \\
\hline
5
&   5.0158 &  2.06827 & -59.5252 \\
& -5.75065 &  2.28982 & -50.2504 \\
& -12.1958 &  5.41985 &  9.34926 \\
&  12.4862 &  6.04493 &  6.64019 \\
&  1.76262 &  10.5456 &  4.97903 \\
&  1.95766 &  10.6191 &  4.97903 \\
& -2.77239 &  12.0802 &  4.97928 \\
&  35.9114 &  27.6808 &  5.00761 \\
& -48.7488 &  32.7442 &  5.06396 \\
\hline\hline
\end{tabular}
\caption{The numerical data of extrema for $F^\text{imp}(2.0,10.0;\alpha_0,m_0^2)$}
\label{table:free10_2}
\end{table}
\renewcommand{\baselinestretch}{1.5}
\renewcommand{\baselinestretch}{0.8}
\begin{table}
\begin{tabular}{cddd}
\hline\hline
\multicolumn{1}{c}{\makebox[5em]{order}} & 
\multicolumn{1}{c}{\makebox[12em]{$\alpha_0$}} & 
\multicolumn{1}{c}{\makebox[12em]{$m_0^{\,2}$}} & 
\multicolumn{1}{c}{\makebox[12em]{$A^\text{imp}$}} \\
\hline
3
& 0.566682 & -0.941477 &  48.9325 \\
& 0.460771 & -0.890109 &  48.8427 \\
& -0.043734 & -0.807144 &   68.646 \\
& 0.113023 & -0.728224 &  58.1744 \\
&  1.92014 &  10.6554 & -0.00337663 \\
& 0.00372349 &  10.8095 & -0.00330928 \\
\hline
4
& -1.33655 & -1.39229 & -116.063 \\
& -0.166586 & -1.07679 &  202.618 \\
& -0.458842 & -0.94533 &   57.275 \\
& 0.131775 & -0.934978 &  108.256 \\
& -0.128625 & -0.778924 & -145.652 \\
& 0.257705 & -0.733339 &   974.75 \\
&  0.12539 & -0.674379 &  885.926 \\
&  1.83575 &  10.5838 & -0.00337685 \\
&  1.90363 &  10.8436 & -0.00337683 \\
& -3.28575 &  12.5122 & -0.00307347 \\
\hline
5
&  1.08531 & -1.82014 &  59.1655 \\
& 0.511521 & -1.44973 &   41.398 \\
& -0.271338 & -1.42093 &  434.787 \\
& 0.174771 & -1.23817 &  73.8638 \\
&  1.90812 &  10.5967 & -0.00337686 \\
&  1.33656 &  10.6194 & -0.0033768 \\
&  1.87274 &  11.0713 & -0.00337686 \\
& -6.75945 &  14.7516 & -0.00278608 \\
\hline\hline
\end{tabular}
\caption{The numerical data of extrema for $A^\text{imp}(2.0,10.0;\alpha_0,m_0^2)$}
\label{table:a10_2}
\end{table}
\renewcommand{\baselinestretch}{1.5}

%================================================================
\end{document}